\documentstyle[preprint,aps]{revtex}
\begin{document}
\draft

\preprint{QMW-PH-96-17}

\title{Linear quantum state diffusion for non-Markovian
open quantum systems}

\author{Walter T. Strunz \footnote{e-mail: W.Strunz@qmw.ac.uk}}

\address{Department of Physics, Queen Mary and Westfield College, University
         of London, Mile End Road, London E1 4NS, United Kingdom}

\date{October 7, 1996}

\maketitle

\begin{abstract}
We demonstrate the relevance of complex Gaussian stochastic 
processes to the stochastic state vector description of non-Markovian
open quantum systems. These processes express the general
Feynman-Vernon path integral propagator for open quantum systems as the
classical ensemble average over stochastic pure state propagators
in a natural way. 
They are the coloured generalization of complex Wiener processes
in quantum state diffusion stochastic Schr\"odinger equations. 

\end{abstract}


\section{Introduction}

The reduced density operator of a quantum subsystem
is obtained from the total density operator by
tracing over the environmental degrees of freedom. 
Feynman and Vernon \cite{FeyVer1} derive 
the propagator ${\cal J}(x_f,x'_f,t;x_0,x'_0,0)$ of the reduced density
matrix $\rho(x,x',t)$ in terms of a 
double path integral [see also Feynman and Hibbs \cite{FeyHib1}]
\begin{equation}\label{propag}
{\cal J}(x_f,x'_f,t;x_0,x'_0,0) =
\int_{x_0,0}^{x_f,t}{\cal D}[x]
\int_{x'_0,0}^{x'_f,t}{\cal D}[x'] \exp\left\{\frac{i}{\hbar}
(S[x] - S[x'])\right\}\;{\cal F}[x,x'],
\end{equation}
where $S[x]$ is the classical action functional of the subsystem alone.
The {\it influence functional} ${\cal F}[x,x']$ 
combines the effects of the environmental initial state, its Hamiltonian and
the interaction Hamiltonian between subsystem and environment, on the 
subsystem.

It is also shown in \cite{FeyVer1,FeyHib1} that the
general influence functional which is at most quadratic in 
the coordinates must be of the form
\begin{equation}\label{inffun}
{\cal F}[x,x'] = \exp\left\{-
\int_0^t d\tau\int_0^\tau d\sigma\; [x_\tau - x'_\tau] 
[\alpha(\tau,\sigma) x_\sigma - \alpha^*(\tau,\sigma) x'_\sigma]
\right\},
\end{equation}
with a positive, Hermitian kernel 
\begin{equation}\label{kerpro}
\alpha(\tau,\sigma) = \alpha^*(\sigma,\tau).
\end{equation}

\subsection{Real kernels}

Feynman and Vernon \cite{FeyVer1,FeyHib1} emphasize that if the kernel is not 
only Hermitian but real, the influence functional 
can be obtained from a real Gaussian stochastic process 
$F(\tau)$ 
(a fluctuating {\it force}), with statistical properties
\begin{eqnarray}\label{stofor}
\langle F(\tau)\rangle & = & 0 \\ \nonumber
\langle F(\tau)F(\sigma)\rangle & = & \alpha(\tau,\sigma) = 
\alpha^*(\tau,\sigma).
\end{eqnarray}
Here, and throughout the paper, $\langle\ldots\rangle$ denotes the classical
ensemble average over the stochastic processes.
Using the general formula
\begin{equation}\label{averea}
\langle\exp
\int_0^t d\tau [f(\tau) F(\tau)] 
\rangle 
 = \exp
\frac{1}{2}
\int_0^t d\tau\int_0^t d\sigma [f(\tau) \alpha(\tau,\sigma) f(\sigma)]
\end{equation}
for arbitrary functions $f(\tau)$,
the propagator ${\cal J}$ for
the density operator can be {\it stochastically decoupled} 
into stochastic pure-state propagators $G_F$,
\begin{equation}\label{unrav}
{\cal J}(x_f,x'_f,t;x_0,x'_0,0) = 
\langle G_F(x_f,t;x_0,0)G^*_F(x'_f,t;x'_0,0)\rangle,
\end{equation}
with 
\begin{equation}\label{patint}
G_F(x_f,t;x_0,0) = 
\int_{x_0,0}^{x_f,t}{\cal D}[x] \exp\left\{\frac{i}{\hbar}
S[x] + i \int_0^t d\tau x_\tau F(\tau) \right\}.
\end{equation}
The total action functional in the exponent of the path integral 
propagator (\ref{patint}) now represents the
additional influence of the stochastic force
$\hbar F(\tau)$ on the subsystem. 
Thus, real kernels are equivalent to ordinary unitary, but stochastic,
quantum dynamics.

\subsection{Complex kernels}

Genuine environments, however, not only induce fluctuations
in the subsystem but also dissipation. These are represented by complex
kernels $\alpha(\tau,\sigma)$
and can therefore not be simulated
by a stochastic potential. As an example, Feynman and Vernon 
\cite{FeyVer1,FeyHib1}
derive the influence functional (\ref{inffun}) analytically
for the case of a linear coupling to a heat bath of
harmonic oscillators with temperature $T$. 
Caldeira and Leggett further elaborate 
this approach in \cite{CalLeg1} [see also Grabert \cite{Graber1} and Weiss
\cite{Weiss1}], resulting in the complex
{\it quantum Brownian motion} kernel
\begin{equation}\label{quabro}
\alpha(\tau,\sigma) = \frac{\gamma m}{\pi\hbar}\int_0^{\Omega} d\omega
\;\omega \left\{\coth\left(\frac{\hbar\omega}{2kT}\right)
\cos\left(\omega(\tau-\sigma)\right) 
  - i \sin\left(\omega(\tau-\sigma)\right) \right\},
\end{equation}
where $\Omega$ is a bath cut-off frequency, $m$ the mass of
the particle and $\gamma$ the damping rate.

Remarkably, it has been shown only recently by Di\'osi \cite{Diosi1} that 
even in the general
case of a complex kernel like (\ref{quabro}), the Feynman-Vernon 
propagator (\ref{propag})
allows a stochastic decoupling similar
to (\ref{unrav}). His result is based on the tricky construction of
a real Gaussian process, whose correlation function
has been given implicitly in terms of $\alpha$. From the point of view
of applications, however,  
the use of these processes appears rather difficult. The 
deeper reason behind 
this construction comes from relativistic measurement theory \cite{Diosi2}.

It is the aim of this paper to present an alternative and much
simpler stochastic
decoupling of the general Feynman-Vernon influence functional, based on
{\it complex} Gaussian stochastic processes. In their white noise
version, these processes have been introduced
from symmetry considerations
in the {\it quantum state diffusion} (QSD) stochastic Schr\"odinger
equation \cite{GisPer0,GisPer1,GisPer2,GisPer3,Strunz1}, describing
Markovian open quantum systems. They also appear in measurement
theories, particularly in cases where the apparatus is represented
by a Bosonic reservoir [see \cite{BelHir95} for more references].

Markovian stochastic state vector methods
have proven indispensable for many applications, particularly
in quantum optics \cite{Carmic1}.
Our result represents a first step
towards an applicable {\it non-Markovian} stochastic state vector theory,
required for instance in solid state theory \cite{Graber1,Weiss1}.

In a recent related work Kleinert and Shabanov \cite{KleSha1},
derive operator quantum Langevin equations 
in the Heisenberg picture corresponding to the propagator (\ref{propag}).
In this paper, however, we stick to path integrals and state vectors
in the Schr\"odinger picture. 
We review basic properties of complex Gaussian processes
in Sect. 2.  In Sect. 3 we show how they enable the
stochastic decoupling of the general Feynman-Vernon propagator, 
resulting in linear
{\it non-Markovian quantum state diffusion}.
We conclude with a short discussion and further comments in the
final Sect. 4.

\section{Complex Gaussian stochastic processes}

Here we review {\it complex} Gaussian processes $Z(\tau)$ with
stochastic properties 
\begin{eqnarray}\label{comflu}
\langle Z(\tau)\rangle & = & 0, \\ \nonumber
\langle Z(\tau)Z(\sigma)\rangle & = & 0\;\;\mbox{and} \\ \nonumber
\langle Z(\tau)Z^*(\sigma)\rangle & = & \gamma(\tau,\sigma).
\end{eqnarray}
Such processes $Z(\tau)$ can only be constructed 
if the complex correlation $\gamma$ is 
Hermitian and positive, which is automatically
fulfilled by the quantum environments we are interested in. 
The processes $Z(\tau)$ with properties (\ref{comflu}) are the natural 
coloured generalization of complex Wiener processes $\xi(\tau)$
with corresponding It\^o increments $d\xi$ with properties
\begin{equation}\label{flupro}
(d\xi)^2 = 0\;\mbox{and}\;|d\xi|^2 = dt,
\end{equation}
as they arise from symmetry considerations 
in the quantum state diffusion theory
of Markovian open quantum systems \cite{GisPer0,GisPer1,GisPer2,GisPer3}. 
Complex processes with properties (\ref{comflu}) are also common in quantum
measurement theories \cite{BelHir95}.

Writing $Z(\tau) = X(\tau) + i Y(\tau)$ we find that conditions 
(\ref{comflu}) are fulfilled for
\begin{eqnarray}\label{reaima}
\langle X(\tau)X(\sigma)\rangle = \langle Y(\tau)Y(\sigma)\rangle & = & 
\frac{1}{2}\mbox{Re}\left\{\gamma(\tau,\sigma)\right\} \;\;\mbox{and} \\ \nonumber
\langle X(\tau)Y(\sigma)\rangle = -\langle Y(\tau)X(\sigma)\rangle & = & 
- \frac{1}{2}\mbox{Im}\left\{\gamma(\tau,\sigma)\right\}. 
\end{eqnarray}
We see that the crucial advantage of complex processes is to 
allow a non-vanishing correlation between their real and imaginary part,
regarding
the real processes $X(\tau)$ and $Y(\tau)$ as one joint real Gaussian
process $(X(\tau),Y(\tau))$. This 
construction leads to the imaginary part of the correlation function 
$\gamma(\tau,\sigma)$ in (\ref{comflu}).

The relevant formula for complex processes $Z(\tau)$ that replaces
formula (\ref{averea}) for real processes is
\begin{equation}\label{comav1}
\langle \exp
\int_0^t d\tau\left[ f(\tau) Z(\tau) + g(\tau) Z^*(\tau) \right] 
\rangle  
 = \exp
\int_0^t d\tau\int_0^t d\sigma\left[ f(\tau) \gamma(\tau,\sigma) g(\sigma)
\right] 
,
\end{equation}
valid for arbitrary functions $f(\tau)$ and $g(\tau)$.
Notice that this implies
\begin{equation}\label{comav2}
\langle\exp
\int_0^t d\tau [f(\tau) Z(\tau)] 
\rangle = 1
\end{equation}
in contrast to equation (\ref{averea}) for real processes.

If one wants to generate such processes $Z(\tau)$ numerically, one
can use the construction
\begin{equation}\label{numeri}
Z(t) = \int d\tau\;\;  \gamma_{\frac{1}{2}}(t,\tau)\xi(\tau),
\end{equation}
where $\xi(\tau)$ is an easily generated white complex process and 
\begin{equation}\label{gamhal}
\int d\tau\;\;\gamma_{\frac{1}{2}}(t,\tau)\gamma_{\frac{1}{2}}(\tau,s) 
= \gamma(t,s).
\end{equation}

\section{Stochastic decoupling of the Feynman-Vernon influence functional}

We show how the processes $Z(\tau)$
naturally lead to a stochastic decoupling of the Feynman-Vernon
influence functional. We choose processes with
the complex conjugate of the Feynman-Vernon kernel from (\ref{inffun}) as
correlation function,
\begin{equation}\label{choker}
\gamma(\tau,\sigma) = \alpha^*(\tau,\sigma).
\end{equation}

According to formula (\ref{comav1}) we find
\begin{eqnarray}\label{firave}
\langle\exp
\int_0^t d\tau\left[
x_\tau Z(\tau) + x'_\tau Z^*(\tau)\right] 
\rangle
& = & \exp
\int_0^t d\tau \int_0^t d\sigma \;\;
[x_\tau \alpha^*(\tau,\sigma) x'_\sigma]
 \\ \nonumber
& = & \exp
\int_0^t d\tau \int_0^\tau d\sigma \;\;
\left[x_\tau \alpha^*(\tau,\sigma) x'_\sigma
+ x'_\tau \alpha(\tau,\sigma) x_\sigma
\right]
,
\end{eqnarray}
where we used (\ref{kerpro}) to get the second line.
This last expression is just the part of the influence
functional (\ref{inffun}) that couples $x$ and $x'$. Notice, however,
that they are decoupled on the left-hand side of equation (\ref{firave}).
We conclude that we can express the propagator of the density matrix
in the decoupled form
\begin{equation}\label{prounr}
{\cal J}(x_f,x'_f,t;x_0,x'_0,0) = 
\langle G_Z(x_f,t;x_0,0)G^*_Z(x'_f,t;x'_0,0)\rangle,
\end{equation}
with the stochastic path integral propagator for state vectors,
\begin{equation}\label{stapro}
G_Z(x_f,t;x_0,0) = 
\int_{x_0,0}^{x_f,t}{\cal D}[x] \exp\left\{\frac{i}{\hbar}
S[x] + \int_0^t d\tau x_\tau Z(\tau)  
-\int_0^td\tau \int_0^\tau d\sigma 
\left[x_\tau\alpha(\tau,\sigma)x_\sigma\right] \right\}.
\end{equation}
Thus, the Feynman-Vernon path integral propagator (\ref{propag}) for
the density operator is equivalent to the ensemble of pure state
propagators (\ref{stapro}).
This is the main result of this paper. The non-local action functional
in (\ref{stapro}) reflects the non-Markovian nature of the problem.

Result (\ref{prounr}) with (\ref{stapro}) allows to describe the 
non-Markovian dynamics of the 
subsystem in terms of an ensemble of stochastic state vectors 
\begin{equation}\label{stopur}
|\psi_Z(t)\rangle = G_Z(t;0)|\psi_0\rangle.
\end{equation}
If we assume an initial
pure state 
\begin{equation}\label{initia}
\rho_0 = P_{\psi_0},
\end{equation}
where we use the notation $P_\psi = |\psi\rangle\langle\psi|$
for pure state projectors,
we recover the density operator at time $t$ by taking the ensemble average
according to (\ref{prounr}),
\begin{equation}\label{denunr}
\rho(t) = \langle P_{\psi_Z(t)}\rangle.
\end{equation}

In Markovian subsystem dynamics \cite{GisPer0,GisPer1,GisPer2,GisPer3},
the time evolution of the stochastic state
vectors is given by a stochastic Schr\"odinger equation.
In fact, in the white noise case
\begin{equation}\label{whinoi}
\alpha(t,s) = \kappa \delta(t-s),
\end{equation}
the path integral propagator (\ref{stapro}) becomes local in time,
\begin{equation}\label{expo}
G_\xi(x_f,t;x_0,0) = 
\int_{x_0,0}^{x_f,t}{\cal D}[x] \exp\left\{\frac{i}{\hbar}
S[x] + \sqrt{\kappa} \int_0^t d\xi(\tau) x_\tau 
-\frac{\kappa}{2} \int_0^td\tau x_\tau^2 \right\},
\end{equation}
with a delta-correlated normalized complex process 
$\xi(t) = \kappa^{-\frac{1}{2}} Z(t)$.
This is the propagator of the linear quantum state
diffusion stochastic Schr\"odinger equation
\begin{equation}\label{LQSD}
|d\psi\rangle = \left(-\frac{i}{\hbar} \hat H - \frac{\kappa}{2}{\hat x}^2\right)
|\psi\rangle dt
+ \sqrt{\kappa}{\hat x}|\psi\rangle d\xi
\end{equation}
with complex It\^o increments (\ref{flupro}). The path integral theory
of general linear quantum state diffusion equations was developed
in \cite{Strunz1}, where a rigorous definition of stochastic path
integrals like (\ref{expo}) is given. 

As the non-Markovian generalization
of (\ref{expo}), the propagator (\ref{stapro}) represents
linear non-Markovian quantum state diffusion.

\section{Conclusions}

We use complex Gaussian stochastic processes to find
a non-Markovian quantum state diffusion theory 
which is equivalent to the Feynman-Vernon density matrix formulation.
Our result offers a helpful tool, since
state vectors are simpler than density operators.
Such a reduction in complexity is essential to tackle many 
realistic problems and is most significant numerically, as
is well recognized in the Markovian case, for instance in quantum optics
\cite{Carmic1}.
To be truly helpful, one must overcome the 
difficulties arising from the fact that the stochastic state vectors
$|\psi_Z\rangle$ of (\ref{stopur}) are not normalized. Also an effective 
algorithm to propagate state vectors
with the non-local path integral (\ref{stapro}) has to be established.

From a quantum foundational point of view
the question arises in what sense the stochastic
state vectors $|\psi_Z\rangle$ can be related to the non-Markovian dynamics 
of individual open quantum systems. 

In the well established Markovian
case, the use of white complex noise arose from symmetry considerations 
in the space of the environment operators of the corresponding master 
equation \cite{GisPer0,GisPer1,GisPer2,GisPer3}. 
For a single, Hermitian environment operator like $\hat x$
in (\ref{LQSD}), however, 
this symmetry argument does not apply. In this paper we have shown
that even in this case, complex Gaussian processes appear 
indispensable as soon as non-Markovian features are taken into
account. These independent arguments underline their relevance
to a stochastic description of general open quantum systems.

\section{Acknowledgment}

I would like to thank Ian C. Percival for helpful discussions and advice.
I am also grateful to Lajos Di\'osi and Todd A. Brun for detailed comments
on the manuscript. This work was made possible by a Feodor Lynen fellowship
of the Alexander von Humboldt foundation.

\end{document}